\documentclass[prl,twocolumn,superscriptaddress,showpacs,aps]{revtex4-1}

\usepackage{color}
\usepackage{graphicx}
\usepackage{bm}
\usepackage{amsmath}

\begin{document}

\newcommand{\Tc}{T_{\text c}}
\newcommand{\Hcii}{H_{\text c2}}
\newcommand{\fq}{\Phi_0}
\newcommand{\lamn}{\lambda_n}
\newcommand{\Gac}{\Gamma_{ac}}
\newcommand{\GVL}{\Gamma_{\text{VL}}}
\newcommand{\GHcii}{\Gamma_{\Hcii}}
\newcommand{\SRO}{Sr$_2$RuO$_4$}
\newcommand{\UPt}{UPt$_3$}

\title{Anisotropy of the Superconducting State in {\SRO}}

\author{C.~Rastovski}
\affiliation{Department of Physics, University of Notre Dame, Notre Dame, Indiana 46556, USA}

\author{C.~D.~Dewhurst}
\affiliation{Institut Laue-Langevin, 6 Rue Jules Horowitz, F-38042 Grenoble, France}

\author{W.~J.~Gannon}
\affiliation{Department of Physics and Astronomy, Northwestern University, Evanston, Illinois 60208 USA}

\author{D.~C.~Peets}
\affiliation{Department of Physics, Graduate School of Science, Kyoto University, Kyoto 606-8502, Japan}
\affiliation{Max Planck Institute for Solid State Research, D-70569 Stuttgart, Germany}

\author{H.~Takatsu}
\affiliation{Department of Physics, Graduate School of Science, Kyoto University, Kyoto 606-8502, Japan}
\affiliation{Department of Physics, Tokyo Metropolitan University, Tokyo 192-0397, Japan}

\author{Y.~Maeno}
\affiliation{Department of Physics, Graduate School of Science, Kyoto University, Kyoto 606-8502, Japan}

\author{M.~Ichioka}
\affiliation{Department of Physics, Okayama University, Okayama 700-8530, Japan}

\author{K.~Machida}
\affiliation{Department of Physics, Okayama University, Okayama 700-8530, Japan}

\author{M.~R.~Eskildsen}
\email{eskildsen@nd.edu}
\affiliation{Department of Physics, University of Notre Dame, Notre Dame, Indiana 46556, USA}

\date{\today}

\begin{abstract}
Despite intense studies the exact nature of the order parameter in superconducting {\SRO} remains unresolved. We have used small-angle neutron scattering to study the vortex lattice in {\SRO} with the field applied close to the basal plane, taking advantage of the transverse magnetization.  We measured the intrinsic superconducting anisotropy between the $c$ axis and the Ru-O basal plane ($\sim 60$), which greatly exceeds the upper critical field anisotropy ($\sim 20$). Our result imposes significant constraints on possible models of triplet pairing in {\SRO} and raises questions concerning the direction of the zero spin projection axis.
\end{abstract}


\maketitle

% Introduction
%%%%%%%%%%%%%%%%%%%%%%%%%
The superconducting state emerges due to the formation and condensation of Cooper pairs, although the exact microscopic mechanism responsible for the pairing in different materials varies and in many cases remains elusive. In the prominent case of strontium ruthenate multiple experimental and theoretical studies provide compelling support for triplet pairing of carriers (electrons and/or holes) and an odd-parity, $p$-wave order parameter symmetry in superconducting {\SRO}~\cite{Mackenzie:03a,Maeno:12a}. At the same time, seemingly contradictory experimental results have left important open questions concerning the detailed structure and coupling of the orbital and spin parts of the order parameter. One example of this predicament is conflicting evidence as to whether the $p$-wave order parameter is chiral~\cite{Sauls:09a,Kallin:12a}.

The motivation for the present work is the unresolved question regarding the anisotropy of the superconducting state of {\SRO}.
The Fermi surface in this material consists of three largely two-dimensional sheets with Fermi velocity anisotropies ranging from 57 to 174~\cite{Mackenzie:03a,Bergemann:03a}, and one would expect an upper critical field ($\Hcii$) anisotropy within this range~\cite{RefWorks:54,Chandrasekhar:93a}. Experiments, however, find a much smaller
$\GHcii = \Hcii^{\perp c}/\Hcii^{\parallel c} \simeq 20$ at low temperature and a near constant
upper critical field when the applied field is within $\pm 2^{\circ}$ of the basal plane~\cite{RefWorks:701}. Within the same angular range the superconducting transition at $\Hcii$ becomes first order, leading to suggestions of a subtle coupling between the magnetic field and the triplet order parameter~\cite{Yonezawa:12a}, or Pauli limiting, which is inconsistent with triplet pairing with the Cooper pair zero spin projection along the $c$ axis~\cite{Machida:08a}.

In this Letter we report on measurements of the intrinsic anisotropy of the superconducting state ($\Gac$) in {\SRO}, which is found to be $\sim 3$ times greater than $\GHcii$. A successful model for the superconducting state in strontium ruthenate must be able to account for the large difference between these two anisotropies.

\begin{figure}
 \includegraphics{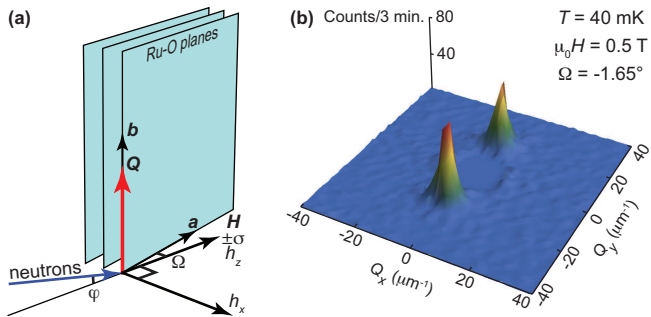}
 \caption{\label{GDP}
          Experimental geometry.
          (a) The coordinate system is defined with $z$ along $\bm{H}$ and $y$ in the Ru-O basal plane (along $\bm{b}$). The applied magnetic field $\bm{H}$ is rotated away from the Ru-O (spanned by $\bm{a}$ and $\bm{b}$) by an angle $\Omega$. Neutron spins ($\sigma$) are parallel or antiparallel to the magnetic field. The incident neutron beam is in the $yz$ plane, at an angle $\varphi$ relative to the field direction. The observed VL scattering vector is denoted $\bm{Q}$ and the longitudinal and transverse component of the field modulation by $h_z$ and $h_x$, respectively.
          (b) Diffraction pattern showing spin-flip scattering from the VL due to the transverse field modulation ($h_x$). The two Bragg peaks correspond to $\pm \bm{Q}$ in panel (a). No background subtraction was performed, but a small remnant of the undiffracted beam close to $Q = 0$ due to the finite flipping ratio ($\sim 8$) is masked off }
\end{figure}

% Experimental details
The anisotropy $\Gac$ was determined by small-angle neutron scattering (SANS) studies of the vortex lattice (VL).
The experiment was performed using a single crystal of {\SRO} grown by the floating zone method and carefully annealed, yielding a critical temperature $\Tc = 1.45$~K and no indication of a 3~K phase~\cite{Mackenzie:03a}. Measurements were performed at $T = 40-60$~mK using a dilution refrigerator inserted into a horizontal-field cryomagnet. Magnetic fields of $\mu_0 H = 0.5$ and $0.7$~T were applied close to the sample $a$ axis. A motorized $\Omega$ stage could rotate the dilution refrigerator within the magnet, allowing {\em in situ} sample alignment and measurements as the crystalline basal plane was rotated with respect to $\bm{H}$. A schematic of the experimental configuration is shown in Fig.~\ref{GDP}(a).
The VL was prepared by changing $H$ and $\Omega$ at the base temperature, followed by a damped small-amplitude field modulation. This method produces a well-ordered VL and eliminates the need for a field-cooling procedure before each measurement.
The SANS experiment was carried out on the D11 and D22 instruments at Institut Laue-Langevin, using a neutron wavelength $\lamn = 1.7$~nm and a wavelength spread $\Delta \lamn/\lamn = 10$\%. Part of the measurements were performed using polarized incident neutrons and a $^3$He analysis cell to allow discrimination between spin-flip and non-spin-flip scattering.

% Transverse VL field modulation
In order to determine $\Gac$ it is necessary to study the VL with the magnetic field oriented parallel or very close to the crystalline basal plane. Such measurements are challenging and require a novel approach to VL SANS studies in order to be feasible. Briefly, the VL scattered intensity is determined by the amplitude of the field modulation and is proportional to $|\bm{h}|^2$, where $\bm{h}(\bm{q})$ is the Fourier transform of the magnetic field $\bm{B}(\bm{r})$~\cite{Eskildsen:11a}. Using state-of-the-art SANS instruments at a high-flux neutron source such as Institut Laue-Langevin, it is possible to measure the diffraction from a well-ordered VL with a longitudinal Fourier coefficient $|h_z|$ as low as $0.1-1$~mT, depending on the amount of background scattering~\cite{Eskildsen:11b}. Here $|h_z| \propto \lambda_{\perp}^{-2}$, where $\lambda_{\perp}$ is the average penetration depth in the screening current plane perpendicular to the applied field. Previous SANS studies with $\bm{H} \parallel \bm{c}$ found a VL form factor for {\SRO} no greater than a few mT~\cite{RefWorks:504}. This indicates that measurements with $\bm{H} \perp \bm{c}$ should not be possible as $|h_z^{\perp c}|^2/|h_z^{\parallel c}|^2 \propto (\lambda_{ab}/\lambda_c)^2 = \Gac^{-2}$, and with $\Gac \geq 20$ we estimate $|h_z^{\perp c}| \leq 3$~$\mu$T, at least 2 orders of magnitude below what is required for a VL SANS experiment. However, in highly anisotropic superconductors such as {\SRO}, there is a strong preference for the vortex screening currents to run within the basal $ab$ plane. A small ``misalignment'' angle $\Omega$ between the applied field and the basal plane will thus lead to a significant transverse Fourier coefficient ($h_x$). Estimates based on an extended London model which includes an effective mass anisotropy yields $|h_x/h_z|^2 \propto \Gac^2$~\cite{RefWorks:794}, and thus predict $h_x^{\perp c}$ to be comparable in magnitude to $h_z^{\parallel c}$. As a result, scattering due to the transverse field modulation should be observable. This is confirmed by the VL diffraction pattern shown in Fig.~\ref{GDP}(b) which shows Bragg peaks aligned with the crystalline $\bm{b}$ direction ($y$ axis).

\begin{figure}
 \includegraphics{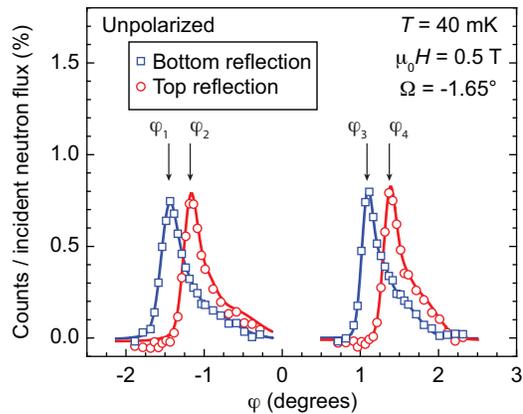}
 \caption{\label{RC}
          Vortex lattice rocking curves showing the scattered intensity as a function of the angle $\varphi$, for an unpolarized neutron beam.
          Error bars are equal to or smaller than the symbols.
          Two maxima are observed for both the bottom ($\varphi_{1/3}$) and top ($\varphi_{2/4}$) VL Bragg reflections (rocking curves obtained with a polarized neutron beam can be found in the Supplemental Material~\cite{SM}).
          The intensity was normalized to the incident neutron flux. The curves are fits to the data as described in the text.}
\end{figure}

% Spin-flip scattering
Scattering from the transverse field modulation leads to a flipping of the neutron spin ($\sigma \perp h_x$) and a Zeeman splitting of the VL rocking curves shown in Fig.~\ref{RC}~\cite{SM}.
Two maxima are observed for both the top [positive $Q_y$ in Fig.~\ref{GDP}(b)] and bottom (negative $Q_y$) VL reflection, as the angle ($\varphi$) between the scattering vector $\bm{Q}$ and the direction of the incident neutron beam is varied to satisfy the Bragg condition. As expected, no scattering from the otherwise more commonly observed longitudinal VL field modulation ($h_z$) could be measured in {\SRO}.  A more detailed discussion of the spin-flip scattering can be found in Ref.~\cite{Kealey:01a}, where a similar but much less extreme effect was observed in yttrium barium copper oxide (YBCO).

To verify that the observed diffraction is due to spin-flip scattering, measurements with a polarized neutron beam were performed (shown in the Supplemental Material~\cite{SM}).
In this case only one maximum is observed for each Bragg reflection, selected according to the direction of the neutron spin. Furthermore, the scattered intensity normalized to the incident neutron flux is doubled relative to the unpolarized beam as expected. Moreover, using polarization analysis it is possible to measure only the spin-flip scattering as shown in Fig.~\ref{GDP}(b).

\begin{figure}
 \includegraphics{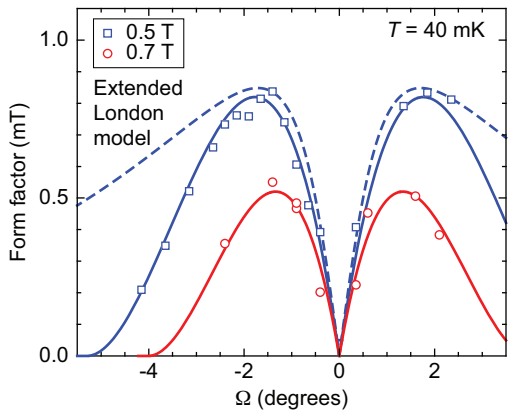}
 \caption{\label{FF}
          Vortex lattice form factor at 40~mK as a function of applied field and angle $\Omega$ with the $a$ axis. The statistical error is roughly the size of the symbols.
          The solid lines are guides to the eye. The dashed line shows an extended London model fit to the $0.5$~T data as discussed in the text, with $\lambda_{ab} = 167$~nm, $\xi_{ab} = 66$~nm, $c =1/4$, and $\Gac = 58.5$.}
\end{figure}

% Measurement window
Dividing the integrated intensity by the incident neutron flux yields the integrated VL reflectivity
\begin{equation}
  R = \frac{2\pi \gamma^2 \lamn^2 t}{16 \fq \, Q} \, |h_x|^2,
\end{equation}
where $\gamma = 1.913$ is the neutron magnetic moment in nuclear magnetons, $t$ is the sample thickness and $\fq = h/2e = 2069$~T$\,$nm$^2$ is the flux quantum~\cite{Eskildsen:11b}. As shown in Fig.~\ref{RC} each peak is fitted to the sum of three Gaussians due to the asymmetry of the rocking curves~\cite{RCasymmetry}. Moreover, the integrated intensity for the two maxima (top, bottom) for a given reflection are added, as each corresponds to half the incident flux (one direction of the neutron spin). The form factor obtained in this fashion is shown in Fig.~\ref{FF}, for all measured fields and $\Omega$'s.

Figure~\ref{FF} %The above discussion
illustrates how the VL SANS measurements are possible within a narrow angular range, with $\bm{H}$ close to, but not perfectly aligned with, the basal plane. The width of the measurement ``window'' decreases with increasing field due to the rapidly decreasing $\Hcii(\Omega)$~\cite{RefWorks:701}. In addition, the overall form factor decreases with increasing field. While the anisotropic London model provides a qualitative description of the enhanced field modulation~\cite{RefWorks:794}, it does not provide a good quantitative fit to the data.
As shown in Fig.~\ref{FF}, an extended London model that includes a so-called core correction by multiplying the calculated $|h_x|$ by $\exp (-c \, Q^2(\Omega) \, \xi_{ab}^2)$ still does not yield a good fit to the data. Here the constant $c$ is of the order unity, $Q(\Omega)$ is the magnitude of the VL scattering vector (see below), and $\xi_{ab} = (\fq/2\pi \Hcii^{\parallel c})^{1/2}$ is the in-plane coherence length~\cite{Eskildsen:11b}.
A quantitatively accurate model for the VL form factor is highly desirable as it would allow a determination of both $\lambda$ and $\xi$.

\begin{figure}
 \includegraphics{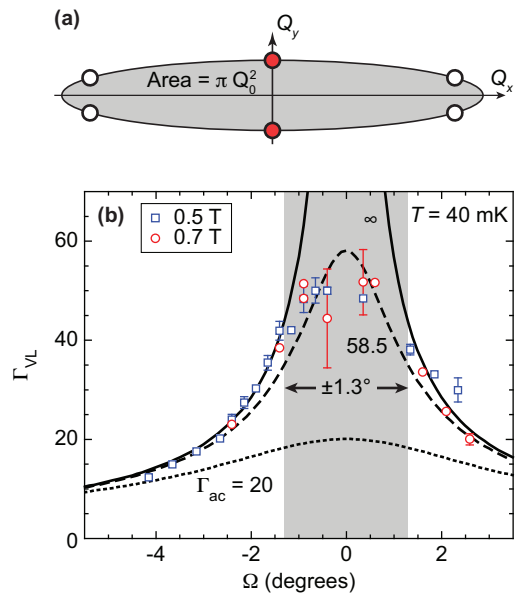}
 \caption{\label{Anisotropy}
          Vortex lattice anisotropy.
          (a) Schematic of VL Bragg reflections lying on an ellipse with major-to-minor axis ratio, $\GVL = 6$. Only the filled (red) peaks are observed. The reciprocal space area of the ellipse is $\pi Q_0^2 = 8\pi^3 \mu_0 H/\sqrt{3}\fq$.
          (b) Measured VL anisotropy at 40~mK as a function of applied field and angle with the $a$ axis ($\Omega$). Except where shown explicitly the statistical error is the size of the symbols. The lines show the VL anisotropy calculated using Eq.~(\ref{Geq}) and $\Gac = 20$ (dotted line), $58.5$ (dashed line), and $\infty$ (full line).}
\end{figure}

% Vortex lattice anisotropy
We now turn to the main result of this Letter, which is the measurement of the VL anisotropy. In an anisotropic superconductor the VL Bragg peaks are expected to lie on an ellipse with a major-to-minor ratio given by~\cite{RefWorks:54}
\begin{equation}
  \GVL = \frac{\Gac}{\sqrt{\cos^2 \Omega + ( \Gac \, \sin \Omega )^2}}
  \label{Geq}
\end{equation}
as shown in Fig.~\ref{Anisotropy}(a).
This $\Omega$ dependence was derived for anisotropic (but still three-dimensional) superconductors, and was verified in early VL SANS measurements on $2H$-NbSe$_2$ with $\Gac = 3.2$~\cite{Gammel:94a}. Although {\SRO} is a layered material, the coherence length along the $c$ axis $\xi_c = 3.3$~nm is still several times greater than the Ru-O interlayer spacing ($0.64$~nm)~\cite{Mackenzie:03a}, and we expect Eq.~(\ref{Geq}) to be applicable~\cite{xidetermination}.

Because of the large anisotropy in {\SRO}, VL Bragg peaks which are not on the vertical axis have scattering vectors essentially parallel to $h_x$, making them unmeasurable as only components of the magnetization perpendicular to $\bm{Q}$ will give rise to scattering~\cite{Squires}. Instead, we determine the VL anisotropy based on flux quantization. Assuming that each vortex carries one flux quantum $\fq$, the area of the reciprocal space ellipse in Fig.~\ref{Anisotropy}(a) is determined uniquely by the applied magnetic field. This yields $\GVL = (Q_0/Q)^2$, where $Q$ is the magnitude of the measured VL scattering vector and $Q_0 = 2\pi (2 \mu_0 H/\sqrt{3}\fq)^{1/2}$ corresponds to an undistorted hexagonal VL ($\Gac = 1$). The magnitude of $Q$ can be determined either from the position of the VL Bragg peaks on the detector as shown in Fig.~\ref{GDP}(b) or from the peak positions $\varphi_1$, \ldots, $\varphi_4$ in Fig.~\ref{RC}~\cite{SM}.
The two methods yield nearly identical results, and using the average $Q$ we obtain $\GVL(\Omega)$ shown in Fig.~\ref{Anisotropy}(b). Within the scatter in the data the results for both fields collapse onto a single curve, increasing upon approaching the $a$ axis and reaching a value slightly higher than 50 before the intensity vanishes.
Theoretical predictions of a field-dependent $\GVL$ and possibly a rotation of the VL are thus not observed~\cite{RefWorks:584,RefWorks:871}.
If one assumes a quantization of $\fq/2$, as recently reported for mesoscopic rings of {\SRO}~\cite{Jang:11a}, the deduced values for $\GVL$ would double. However, we consider this an unrealistic scenario in the present case, with a macroscopic, homogenous sample.

Fitting the data in Fig.~\ref{Anisotropy}(b) to Eq.~(\ref{Geq}) yields $\Gac = 58.5 \pm 2.3$. Only for angles within $\pm 1.3^{\circ}$ does the measured anisotropy deviate from that expected for an infinite $ac$ anisotropy. Also shown for comparison is $\GVL$ expected from the low temperature $\GHcii = 20$~\cite{RefWorks:701} and which provides a very poor fit to the data. We note that $\GHcii$ increases with temperature and reaches a value of $\sim 60 \approx \Gac$ at $\Tc$~\cite{Kittaka:09b}. In addition, the fitted value of $\Gac$ coincides with the anisotropy of the $\beta$ Fermi surface sheet (57)~\cite{Mackenzie:03a,Bergemann:03a}.

% Discussion
The large difference between $\GHcii$ and the intrinsic anisotropy of the superconducting state deep within the mixed phase measured by $\Gac$ indicates a strong suppression of the upper critical field in {\SRO} for $\bm{H} \perp \bm{c}$. One possible explanation for this difference is Pauli limiting due to the Zeeman splitting of spin-up and spin-down carrier states by the applied magnetic field and the resulting reduction of the superconducting condensation energy~\cite{RefWorks:811}. In spin-triplet superconductors the order parameter is most conveniently described in terms of the $d$ vector, directed along the zero spin projection axis where the configuration of the Cooper pairs is given by $\tfrac{1}{\sqrt{2}} (\left|\uparrow\downarrow\right\rangle + \left|\downarrow\uparrow\right\rangle)$~\cite{Mackenzie:03a,Maeno:12a,Kallin:12a}. Consequently, Pauli limiting in the triplet case can only occur when $\bm{H} \parallel \bm{d}$.
If one assumes Pauli limiting our results are thus inconsistent with the chiral superconducting state with $\bm{d} \parallel \bm{c}$ proposed for {\SRO}~\cite{Maeno:12a,Kallin:12a}. It should be noted, however, that Pauli limiting itself appears to be in disagreement with Nuclear Magnetic Resonance and Nuclear Quadrupole Resonance Knight-shift measurements (summarized in Ref.~\cite{Maeno:12a}), which suggest that the $d$ vector rotates in the presence of a magnetic field such that $\bm{d} \perp \bm{H}$.

Also remaining are a number of other models for the superconducting state in strontium ruthenate which are (or may be) consistent with our results. Among these are several possible ways to achieve a subtle coupling between the magnetic field and the triplet order parameter as discussed in some detail in Ref.~\onlinecite{Yonezawa:12a}.
Other alternatives include (chiral) triplet pairing with $\bm{d} \perp \bm{c}$~\cite{Miyake:2010ej} that could possibly be locked along certain in-plane directions, recent multiband $p$-wave models~\cite{Chung:12a}, a field-dependent mixing of singlet and triplet states~\cite{Puetter:12a}, or singlet superconductivity~\cite{Zutic:05,Machida:08a}. It should be noted, however, that $s$-wave superconductivity does not provide a satisfactory explanation for the extreme sensitivity of $\Tc$ to impurities or to the chiral properties of {\SRO}~\cite{Mackenzie:03a,Zutic:05,Maeno:12a}. Further experimental and theoretical work will be necessary to provide a definitive determination of the order parameter in this material.

% Conclusion
In conclusion, we have used SANS to measure the anisotropy of the superconducting state in {\SRO}, taking advantage of the transverse VL field modulation which allows measurements in a narrow range of field angles close to, but not perfectly aligned with, the Ru-O basal plane. The superconducting anisotropy greatly exceeds that of the upper critical field and imposes significant constraints on the possible pairing of carriers in this material. Any model aimed at describing the superconducting phase must provide a satisfactory explanation for this observation.

% Acknowledgments
We acknowledge discussions with W. P. Halperin, V. G. Kogan, I. Mazin, J. A. Sauls and S. Yonezawa, and assistance with sample alignment by G. Sigmon. Research support was provided by the U.S. Department of Energy, Office of Basic Energy Sciences, under Award No.~DE-FG02-10ER46783 (neutron scattering) and by the MEXT of Japan KAKENHI No. 22103002 (crystal growth and characterization).


\begin{thebibliography}{99}
\bibitem{RCasymmetry}
The reason for the rocking curves asymmetry is a field inhomegeneity, which is a well-known problem with the particular cryomagnet used for the experiment. The asymmetry does not affect the analysis or conclusions of this Letter.
\bibitem{xidetermination}
The value for $\xi _c$ is obtained from the upper critical field assuming orbital limiting, $H_{c2}^{\perp c}=\Phi_0/2\pi\xi_{ab}\xi_c$. However, even with substantial Pauli limiting $\xi_c$ will be greater than the Ru-O interplane spacing. For the two distances to be equal would require $H_{c2}^{\perp c}\approx 7.7$~T, more than 5 times greater than the measured upper critical field.
\end{thebibliography}
\end{document}

\newcommand{\Tc}{T_{\text c}}
\newcommand{\Hcii}{H_{\text c2}}
\newcommand{\fq}{\Phi_0}
\newcommand{\lamn}{\lambda_n}
\newcommand{\Gac}{\Gamma_{ac}}
\newcommand{\GVL}{\Gamma_{\text{VL}}}
\newcommand{\GHcii}{\Gamma_{\Hcii}}
\newcommand{\SRO}{Sr$_2$RuO$_4$}
\newcommand{\UPt}{UPt$_3$}

\title{Supplementary Material: Anisotropy of the superconducting state in {\SRO}}

\author{C.~Rastovski}
\affiliation{Department of Physics, University of Notre Dame, Notre Dame, Indiana 46556, USA}

\author{C.~D.~Dewhurst}
\affiliation{Institut Laue-Langevin, 6 Rue Jules Horowitz, F-38042 Grenoble, France}

\author{W.~J.~Gannon}
\affiliation{Department of Physics and Astronomy, Northwestern University, Evanston, Illinois 60208 USA}

\author{D.~C.~Peets}
\affiliation{Department of Physics, Graduate School of Science, Kyoto University, Kyoto 606-8502, Japan}
\affiliation{Max Planck Institute for Solid State Research, D-70569 Stuttgart, Germany}

\author{H.~Takatsu}
\affiliation{Department of Physics, Graduate School of Science, Kyoto University, Kyoto 606-8502, Japan}
\affiliation{Department of Physics, Tokyo Metropolitan University, Tokyo 192-0397, Japan}

\author{Y.~Maeno}
\affiliation{Department of Physics, Graduate School of Science, Kyoto University, Kyoto 606-8502, Japan}

\author{M.~Ichioka}
\affiliation{Department of Physics, Okayama University, Okayama 700-8530, Japan}

\author{K.~Machida}
\affiliation{Department of Physics, Okayama University, Okayama 700-8530, Japan}

\author{M.~R.~Eskildsen}
\email{eskildsen@nd.edu}
\affiliation{Department of Physics, University of Notre Dame, Notre Dame, Indiana 46556, USA}

\date{\today}

\maketitle

The two different directions of the neutron spin with respect to the applied field correspond to different nuclear Zeeman energies and lead to opposite shifts of the neutron momentum vector
\begin{equation}
  k_{\uparrow(\downarrow)} = k_0 \, \sqrt{1 \pm \Delta \varepsilon/\varepsilon_0},
\end{equation}
where the subscript in parentheses henceforth corresponds to the lower (in this case minus) sign in the $\pm \Delta \varepsilon$ term. Here the nominal neutron wavevector $k_0 = 2\pi/\lamn$, $\varepsilon_0 = \hbar^2 k_0^2/2m_n$, $\Delta \varepsilon = \gamma \mu_N B$ and $\gamma = 1.913$ is the neutron magnetic moment in nuclear magnetons $\mu_N = e\hbar/2m_n = 31.5$~neV/T. With $\lamn = 1.7$~nm and $\mu_0 H = 0.5$~T one finds $k_{\uparrow}^2 - k_{\downarrow}^2 = 2 \times 10^{-4} \, k_0^2$.
\begin{figure}[b]
\includegraphics{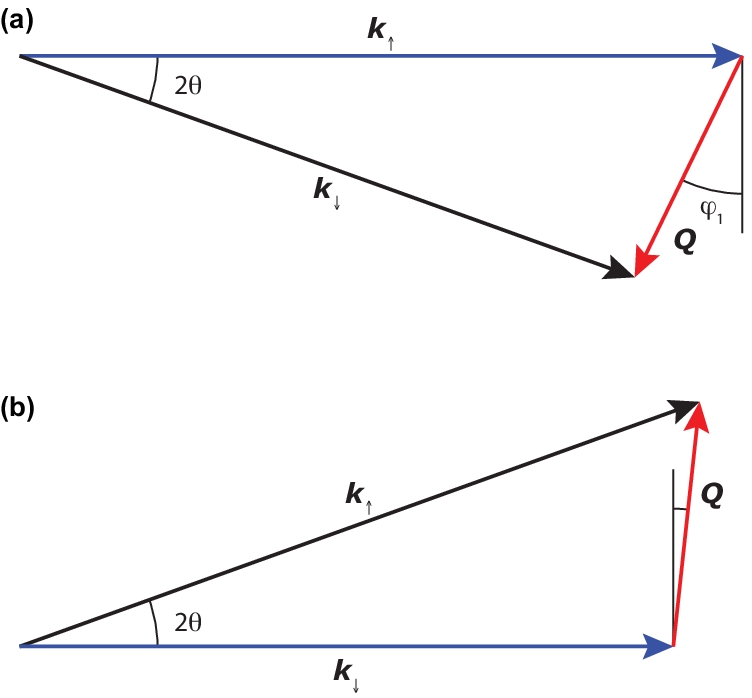}
\caption{\label{Bragg}
          Schematics showing the scattering geometries corresponding to the reflection at $\varphi = \varphi_1$ (a) and $\varphi_2$ (b).
          The colors of the incident neutron wavevector (blue) and the scattering vector (red) correspond to those used in Fig.~1 of the main text.}
\end{figure}

Due to the short $Q$ in the range $0.003 - 0.01 \, k_0$, the small shift in the neutron wavevector nonetheless leads to a significant difference in the angle ($\varphi_{1(2)})$ required to satisfy the Bragg condition as shown schematically in Fig.~\ref{Bragg}.
In this case Bragg's law is replaced by
\begin{equation}
  k_{\uparrow}^2 - k_{\downarrow}^2 \pm Q^2 = 2k_{\uparrow(\downarrow)} \, Q \sin \varphi_{1(2)}.
\end{equation}
%In contrast, the scattering angles $\theta_1/\theta_2 = k_{\downarrow}/k_{\uparrow}$ are essentially identical for the two scattering configurations. %% THIS IS NOT TRUE
The magnitude of $Q$ can be determined from the peak positions $\varphi_1$, \ldots, $\varphi_4$ by
\begin{eqnarray}
  Q_{1(2)} & = & \mp k_{\uparrow(\downarrow)} \, \sin \varphi_{1(2)} \mp \sqrt{k_{\downarrow(\uparrow)}^2 - k_{\uparrow(\downarrow)}^2 \, \cos^2 \varphi_{1(2)}} \\
  Q_{3(4)} & = & \mp k_{\downarrow(\uparrow)} \sin \varphi_{3(4)} \pm \sqrt{k_{\uparrow(\downarrow)}^2 - k_{\downarrow(\uparrow)}^2 \, \cos^2 \varphi_{3(4)}}.
\end{eqnarray}

\begin{figure}[h]
 \includegraphics{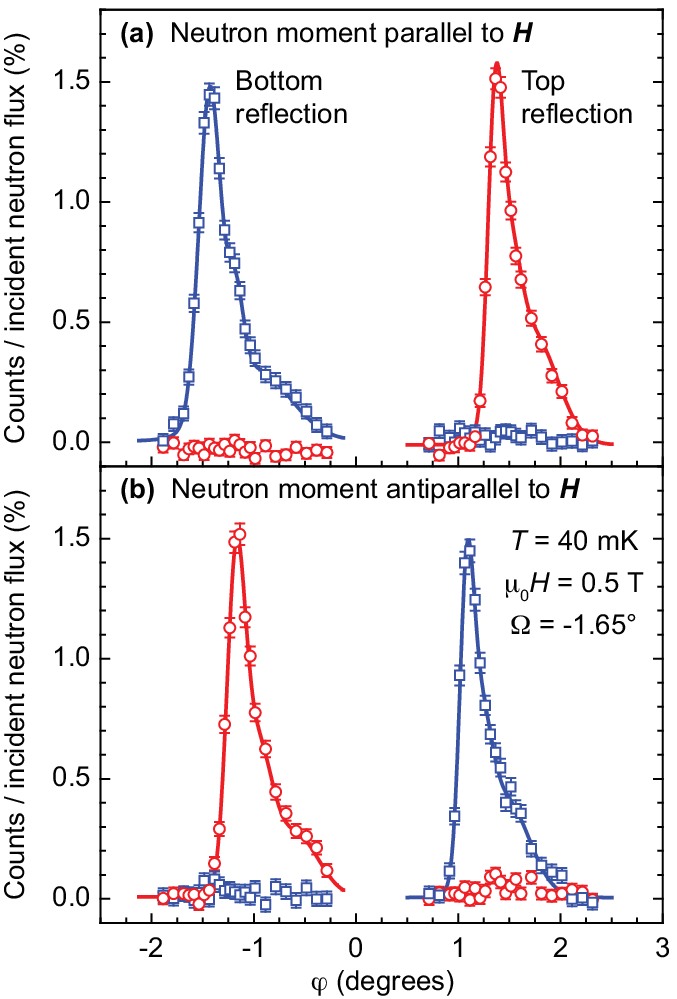}
 \caption{\label{RCpol}
          Vortex lattice rocking curves showing the scattered intensity as a function of the angle $\varphi$, for neutrons polarized with their magnetic moment parallel (a) and antiparallel (b) to the applied field. Except where shown, error bars are equal to or smaller than the symbols. The intensity was normalized to the incident neutron flux. The curves are fits to the data as described in the main text.}
\end{figure}

Rocking curves obtained using a polarized neutron beam are shown in Fig.~\ref{RCpol}. These demonstrate how a single peak can be selected for each Bragg reflection according to the direction of the neutron spin.
